\documentclass{mc-abstract}

\usepackage{graphicx}
\usepackage{enumitem}
\usepackage{csquotes}
\usepackage{quoting} 

\usepackage{multirow}

\usepackage{makecell}

\usepackage{array}

\usepackage{hyperref} %

\usepackage{amssymb}
\usepackage{pifont}

\usepackage{makecell}

\usepackage{caption} 
\year{2023}

\title{Tools for Refactoring to Microservices: \\A Preliminary Usability Report}
\author{Jonas Fritzsch$^1$ \and Filipe Correia$^2$ \and Justus Bogner$^3$ \and Stefan Wagner$^1$}
\affiliation{
    $^1$ Institute of Software Engineering, University of Stuttgart, Germany \\ 
    $^2$ Faculty of Engineering, University of Porto, Portugal \\
    $^3$ Department of Computer Science, Vrije Universiteit Amsterdam, The Netherlands
}
\correspondence{jonas.fritzsch@iste.uni-stuttgart.de}

\begin{document}

\maketitle

\abstract{
While Microservices are a preferred choice for modern cloud-based applications, the migration and architectural refactoring of existing legacy systems is still a major challenge in industry.  
To address this, academia has proposed many strategies and approaches that aim to automate the process of decomposing a monolith into functional units.
In this study, we review existing migration approaches regarding techniques used and tool support.
From 91 publications, we extracted 22 tools, 7 of which address service decomposition. 
To assess them from an end-user perspective, we investigated their underlying techniques, installation, documentation, usability and support. 
For 5 of them, we generated service cuts using reference applications. 
The results of our preliminary work suggest that the inspected tools pursue promising concepts, but lack maturity and generalizability for reliable use by industry. 
}

\section{Introduction}
The microservices architectural style has been established to date as a \textit{de facto} standard for modern cloud-based software systems.
However, the adoption of Microservices and migration of an existing system can be very challenging~\cite{Fritzsch2019d}. 
While a variety of design guidelines, pattern languages, and best practices are available for greenfield developments~\cite{newman2021,vale2022}, the process of migrating an existing system is not easily generalized. 
To this end, various methods and approaches have been suggested that aim to systematize and automate this process~\cite{Fritzsch2019a,Bajaj2021,Abdellatif2021a,Kirby2021}. 
However, their applicability heavily depends on several aspects of the system, e.g., the availability and preparation of input artifacts, targeted quality attributes, or the maturity of accompanying tools ~\cite{Fritzsch2022b}. 

Practitioners searching for such approaches and tools in scientific publications rarely find them examined from an end-user perspective. While interviewing practitioners, we even found that they are often unknown to them~\cite{Fritzsch2019d}.
Secondary studies offer some perspective, but fall short of providing practical insights gained by testing such tools in detail.
Hence, in this study, our goal is to investigate the current state of research on automating architectural refactoring to microservices regarding techniques and tools, from the viewpoint of potential users -- software architects and developers.
We therefore review academic literature to summarize the current state of research on techniques and tool support for automating the decomposition into services. We install existing tools, and perform decompositions, while documenting our experiences with their underlying techniques, installation, documentation, usability, and support.

\section{Background and Related Work}
Microservices address the complexity of a system through functional decomposition.
However, when transforming an existing monolithic system, developers need reliable splitting criteria. 
Teams often decompose systems by business capabilities or subdomains, e.g., through Domain-Driven Design~\cite{evans2004}, but such techniques still rely a lot on intuition.
Researchers have therefore proposed a variety of automatic and semi-automatic approaches~\cite{Stojkov2021}, which commonly rely on static analysis techniques, sometimes combined with dynamic analysis~\cite{Andrade2022}.
Schröer et al.~\cite{Schroer2020} contributed a large review of 31 approaches. The majority relies on static analysis of the monolith (17), followed by techniques relying on requirements or models as input.
Dynamic analysis, i.e., observing and analyzing the monolith during runtime, was used by only four approaches. 
Bajaj et al.~\cite{Bajaj2021} similarly reviewed 21 approaches, including tool support for automation.
However, most of the reported tools are third-party static or dynamic analysis tools that do not generate service cuts.
Kirby et al.~\cite{Kirby2021} reviewed 13 approaches and collected expectations from 10 practitioners regarding tools that automate the decomposition.

The results indicate that such tools should ideally allow suggesting and comparing decompositions using multiple relationship types simultaneously.
Lapuz et al.~\cite{Lapuz2021} focused on dynamic tooling for service identification. 
Four of the eight analyzed tools address the monolith to Microservices migration. 
The most comprehensive review by Abdellatif et al.~\cite{Abdellatif2021a} covers 41 service identification approaches, from which 13 target Microservices.

\begin{table*}[h!]
    \fontsize{8.5}{11}\selectfont
    \centering
	\caption{Analysis Technique Types in Existing Reviews}
	\label{table:AnalysisTechniqueTypes}
	\begin{tabular}
        {
            >{\raggedright\arraybackslash}p{0.2\textwidth}
            >{\centering\arraybackslash}p{0.065\textwidth}
	    >{\raggedright\arraybackslash}p{0.65\textwidth}
         }
	Review (Year) & Sources & Types of Analysis Techniques\\ 
        \hline
        \hline
            Schröer~et~al.~\cite{Schroer2020}~(2020) & 31 & Monolith \textbf{static} (17), Greenfield (9), Business Process (5), Monolith \textbf{dynamic} (4), Monolith first (2)\\
            \hline
            Bajaj et al. \cite{Bajaj2021} (2021) & 21 & \textbf{Static:} Requirement Docs/Models (8), Design Docs (4), Source Code (7), VCS History (2), \textbf{Dynamic} (7), Hybrid (3)\\
            \hline
            Kirby et al. \cite{Kirby2021} (2021) & 13 & Structural-\textbf{static} (11), Structural-\textbf{dynamic} (6), Semantic (2), Evolutionary (1)\\
            \hline
            Lapuz et al. \cite{Lapuz2021} (2021) & 8 & \textbf{Dynamic} analysis (8)\\
            \hline
            Abdellatif~et~al.~\cite{Abdellatif2021a}~(2021) & 13 of 41 & \textbf{Static} analysis (10), \textbf{Dynamic} analysis (5), Lexical analysis (2)\\
            \hline
		\hline
	\end{tabular}
\end{table*}

\noindent
The reviews (Table~\ref{table:AnalysisTechniqueTypes}) reveal a dominance of static analysis approaches operating on models, requirements, and source code.
Dynamic analysis is less common and mostly complements the former technique. 
Some reviews~\cite{Bajaj2021,Abdellatif2021a,Lapuz2021} also cover tools like \textit{Structure101, JPROF, SonarGraph Architect, DBeaver} for static analysis or \textit{Kieker, ExplorViz, JProﬁler} for dynamic analysis.
However, while they cover the analysis part, they do not generate decompositions. 

We are not aware of existing studies inspecting such tools from a user perspective to understand how well they are prepared for industry adoption.
In this study, we seek to shed light on this aspect by focussing on tools for automating the decomposition task.

\section{Method}

We employed the method of a \textit{Rapid Review}~\cite{Cartaxo2020}, which represents a lightweight literature review to collect evidence in a timely manner.
Our guiding research question was the following: 
\vspace{-3pt}
\begin{itemize}[label={RQ:},leftmargin=*]
    \item \textit{How practically usable are existing tools for automating the refactoring to Microservices?} 
\end{itemize}
\vspace{-3pt}
We queried the scientific databases and search engines ACM Digital Library, IEEE Xplore, Springer Link, and Google Scholar using the following generic search string:

\vspace{-2pt}
{
\fontsize{9}{10}\selectfont
\ttfamily
\centering
\begin{verbatim} 
             ("micro[-]service*") [AND "monolith*"] [AND ("refactor*" OR "transform*" OR
            "migrat*" OR "decompos*" OR "partition*" OR "adopt*" OR "tool*" OR "utilit*")]
\end{verbatim}
}
\vspace{-2pt}

\begin{figure}[ht]
    \centering
    \includegraphics[keepaspectratio=true, width=0.6\textwidth]{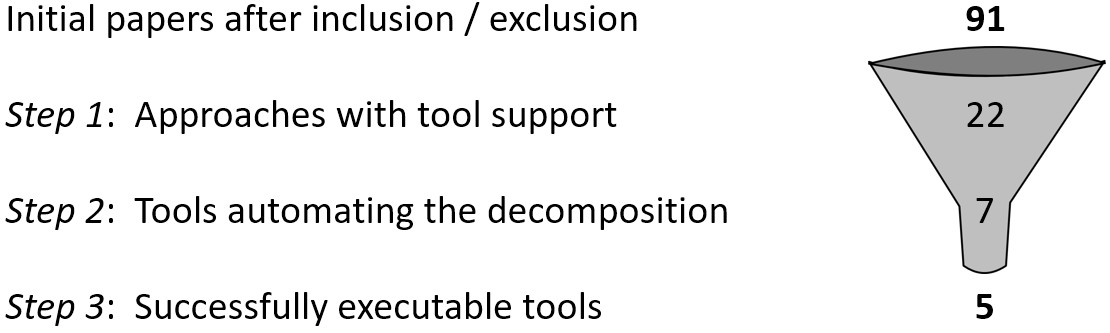}
    \caption{Rapid Review Process: Filtering for Tools Addressing the Decomposition}
    \label{fig:ReviewFilter}
\end{figure}

\noindent
The resulting publications were filtered based on the following inclusion criteria: 1) The paper must describe an approach to migrate a monolithic system to a Microservices-based architecture and 2) The paper must be peer-reviewed, written in English and published in 2016 or later. 
Afterward, we performed backward and forward snowballing until no new work was discovered.
Figure \ref{fig:ReviewFilter} shows that 91 initial papers were identified with this process.

\vspace{2mm}

\noindent
Based on this literature corpus, we applied subsequent filtering to identify tools addressing the decomposition.
\textit{Step 1} yielded 22 papers that report on tools for automating a part of the migration process. 
Auxiliary tools, e.g., tools for solely creating graphical representations, have been excluded in this step already.
In \textit{Step 2}, we filtered further for tools that specifically address the decomposition task.
The remaining seven approaches and tools have been inspected in-depth.
In \textit{Step 3}, we acquired these tools and installed them according to their documentation on a dedicated system. 
This activity resulted in five successfully executable tools (see Table \ref{table:ExecutableTools}) that we set up for testing with suitable reference applications. 
In the 91 sources, we identified several such, mostly Java-based, applications that have been used frequently for evaluations and comparisons, e.g., \textit{JPetStore} (7), \textit{AcmeAir} (6), \textit{DayTrader} (6), and \textit{Cargo Tracking System} (5).
We chose them according to their suitability for the tools' input data format.
To limit efforts, we did not use the same application for all tools, also because the result comparison was not our intended focus.
Each tool was inspected and assessed regarding installation, documentation, usability, and support.
Three software engineering master's students separately inspected the tools, discussed their experiences, and formed a consensus using a three-point ordinal scale (A-C), with A reflecting a good, B a medium, and C a poor experience.
The details of our rapid review and tool inspections can be found online.\footnote{\url{https://doi.org/10.5281/zenodo.7949154}}

\section{Results}

We briefly introduce and characterize the five tools that we were able to execute (see Table \ref{table:ExecutableTools}).
In this regard, we take the practitioner's perspective and share our experiences on their \textit{installation}, \textit{documentation}, \textit{usability}, and \textit{support}.

The tool \textbf{Mono2Micro} contributed by the GitHub organization \textit{socialsoftware} was published in 2019 under the MIT License and is currently under active development \cite{Andrade2022}. 
The creators' goal was to provide an automated workflow for the identification of Microservices considering the trade-off between consistency and partition tolerance in the sense of the \textit{CAP theorem}.
The tool's usability profits from a web-based UI visualizing the decomposition as a graph. The interface allows for several adjustments, even if functionalities are not always easy to locate and interpret.
The developers provided support by promptly answering our questions and handling our submitted GitHub issue.

The tool \textbf{Service Cutter} was contributed by Gysel et al. \cite{Gysel2016} in 2016 and is freely available under the Apache License 2.0 license. 
The web-based tool follows a model-driven approach and thereby achieves a high degree of general applicability.
While Service Cutter's usability profits from the clean UI and its wide applicability, generating the necessary input for industry-scale applications is not feasible without additional tooling.
Service Cutter is no longer maintained as of March 2021, and hence developer support cannot be expected.

\vspace{-2pt}
\begin{table*}[h!]
    \fontsize{8.5}{11}\selectfont
    \centering
	\caption{Characterization of Executable Tools for Decomposition}
	\label{table:ExecutableTools}
	\begin{tabular}
        {
	    p{0.02\textwidth}
	    >{\raggedright\arraybackslash}p{0.26\textwidth}
	    >{\raggedright\arraybackslash}p{0.15\textwidth}
            >{\centering\arraybackslash}p{0.10\textwidth}
            >{\centering\arraybackslash}p{0.11\textwidth}
            >{\centering\arraybackslash}p{0.10\textwidth}
            >{\centering\arraybackslash}p{0.07\textwidth}
	}
	ID & Tool Name & Technique & Installation & Documentation & Usability & Support\\ 
        \hline
        \hline
		1 & Mono2Micro$^a$ (socialsoftware) & static code & B & - & B & B\\
		\hline
            2 & Service Cutter$^b$ & static model & A & B & B & C\\
            \hline
 		3 & Mono2Micro$^c$ (IBM) & static \& dynamic & B & A-B & B & A\\
		\hline
            4 & Microservices Identification$^d$ & static lexical & A & - & C & C\\
            \hline
            5 & MonoBreaker$^e$ & static \& dynamic & B & B-C & B & B-C\\
            \hline
		\hline
	\end{tabular}
         \begin{flushleft}
            \fontsize{8.5}{10}\selectfont
            A = easy/good, B = fair/sufficient, C = poor/insufficient\\ 
            \vspace{2mm}
            \fontsize{8}{10}\selectfont
            $^a$ \url{https://github.com/socialsoftware/mono2micro},\\
            $^b$ \url{https://servicecutter.github.io},\\
            $^c$ \url{https://www.ibm.com/cloud/mono2micro},\\
            $^d$ \url{https://github.com/miguelfbrito/microservice-identification},\\
            $^e$ \url{https://github.com/tiagoCMatias/monoBreaker}
        \end{flushleft}
\end{table*}
\vspace{-2pt}

IBM offers \textbf{Mono2Micro}, an AI-based and semi-automated tool for refactoring monolithic Java applications into Microservices. 
The tool is under active development and can be used under a paid license with optional professional support.
Mono2Micro is an industry-ready tool with solid command-line usability and combines static and dynamic analysis for optimal results.
However, it can only be used for applications running on Websphere Application Server Liberty or Open Liberty.

The tool \textbf{Microservices Identification} was published in 2019 on GitHub by Brito~\cite{Brito2021} and aims to identify plausible decompositions of Java-based monoliths into Microservices.
For that, the tool relies on a semantic technique based on lexical analysis.
The documentation is brief: calculated metrics as well as command lines options are not sufficiently explained.
Likewise, the solely textual output with no functionality to comfortably inspect the resulting service cuts negatively affects usability.
Developer support cannot be expected as the project is no longer maintained.

The tool \textbf{MonoBreaker} by Matias et al.~\cite{matias2020} is the second one that combines static and dynamic analysis.
It is available under the MIT license, and can operate on applications developed using Python with the \textit{Django web framework} and the \textit{Django REST framework}. 
This limits its applicability but makes for a refreshing exception to the prevalent Java focus of the other analyzed tools.
Our usability experiences are based on running the tool over the command line in an industrial context and revealed a smooth installation and use. The results, however, cannot be visually inspected.
Developer support can not be expected either, as the project is no longer maintained.

\section{Limitations}

The rapid review method bears the risk of missing out on important publications.
We tried to mitigate this by relying on existing reviews and performing exhaustive snowballing.
The selection of tools for inspection was influenced by the researchers' capabilities in successfully installing and running them. 
The generalizability of our tool inspection is limited, as it mainly conveys the experiences made by software engineering master's students. We tried to mitigate this threat and minimize subjectiveness of the ratings through discussions in the group and involving experienced researchers in the field.
We chose sample applications which ensured the tools' input requirements to be fully met. Our usability ratings have to be interpreted in this regard.
Moreover, we did explicitly not assess the decomposition results of these tools, i.e., the plausibility and soundness of the generated service cuts.

\section{Discussion and Conclusion}

In this work, we reviewed the tool support for Microservices migrations, targeting the decomposition task. 
From initially 91 approaches, we filtered out five executable and practically usable tools.
All of them use static analysis techniques, while only IBM's \textit{Mono2Micro} and \textit{MonoBreaker} apply dynamic analysis of the monolith in addition.
Three out of five tools are limited to Java-based source code, one requires Python input, while \textit{Service Cutter} is the only language-agnostic tool.
When estimating their technological maturity on the scale of technology readiness levels~\cite{Mankins1995}, we locate the inspected tools between 4 and 6 with IBM's \textit{Mono2Micro} being the most mature tool. 
Despite the outlined limitations, our work can provide guidance to professionals seeking a tool for adoption.
We encourage tool builders from inside and outside academia to follow up on these promising examples. 
For that, the numerous existing approaches we reviewed can serve as a valuable groundwork to build upon.
Essential for practical applicability are generalizability, meaningful visualizations, and a sound documentation, ideally accompanied by a use case example. 
We will continue our efforts to thoroughly evaluate tools for automating the decomposition into Microservices from the practitioners' perspective.
As important future work, we regard comparatively assessing the quality of generated service cuts, particularly involving industry-scale systems.
Involving grey literature might bring to light additional approaches and tools not covered by academia.

\bibliographystyle{plain}
\bibliography{biblio.bib}

\end{document}